\documentstyle[12pt]{article}

\begin{document}
\baselineskip 5mm

\begin{center}{\Large \bf Entropy Revisited: The Plausible Role of 
Gravitation}
\end{center}

\medskip
\medskip

\begin{center}{\bf Eshel Ben-Jacob$^{(a)}$, Ziv Hermon$^{(a)}$, 
Alexander Shnirman$^{(a),(b)}$}
\end{center}

\begin{center}{$^{(a)}$School of Physics and Astronomy,\\
Tel Aviv University, Tel Aviv 69978, Israel\\}
\end{center}

\begin{center}{$^{(b)}$Department of Physics, 
University of Illinois at Urbana-Champaign,\\ 
1110 West Green Street, 
Urbana, Illinois 61801-3080, U.S.A.}
\end{center} 

\begin{abstract}
We first present open questions related to the foundations of 
thermodynamics and statistical physics. We then argue that in 
principle one can not have ``closed systems'', and that a universal 
background should exist. We propose that the gravitational field 
plays this role, due to its vanishing energy-momentum tensor. This 
leads to a new possible picture, 
in which entropy and irreversibility in macroscopic systems emerge 
from their coupling to the background gravitational field.
\end{abstract}

Thermodynamics and statistical physics are the scientific disciplines 
devoted to the description of macroscopic systems at equilibrium. 
Quoting Callen \cite{Callen}: ``whether we are physicists, chemists, 
biologists, or engineers, our primary interface with nature is through 
the properties of macroscopic matter''. Both disciplines are 
considered to be well established. Yet they pose open 
fundamental questions, or even paradoxes. The 
origin of time irreversibility of macroscopic systems is one
example. So are the origin of entropy as a real macroscopic variable, 
the fundamental relation $S = S(U,V,N)$ ($U$ being the system's 
internal energy) and the second law. However, at present many 
physicists think that these are not ``real'' questions. 
The reductionist approach that dominates physical thought
\cite{Weinberg,Mayr} regards the need for the additional laws of 
thermodynamics (on top of the microscopic laws) and the additional 
fundamental assumptions of statistical physics as a 
reflection of our intellectual limitations and not as an ontological 
reality. In principle, they are believed to be derivable from the 
microscopic laws.

The first goal of this paper is to convince the reader that there are
open questions in the foundations of thermodynamics and
statistical physics. The second is to propose that the answers to 
the above questions might have to do with the special nature of 
gravitation, i.~e., the vanishing of the total energy momentum tensor 
of gravitation and matter \cite{Dirac,LL}. Although this property by 
itself seems to be in contradiction with both statistical physics and 
quantum mechanics, we present here a new picture synergizing the above 
disciplines. Over the years there has been considerable effort to 
reconcile classical thermodynamics with classical and quantum gravity.
\cite {Therm_Grav}. In this paper we discuss
the idea that, as every system is coupled to the gravitational metric,
no system can be truly isolated. We will argue that this is the 
source of irreversibility in Nature. This can be shown by tracing over 
the degrees of freedom of the gravitational metric. The 
system plus the metric become then a reduced density matrix, describing
the system only, and this reduced density matrix evolves in the
usual manner towards its final equilibrium state.

Thermodynamics is actually a summary of experimental observations of
properties and of quasistatic processes in macroscopic systems
\cite{Callen}. They all fit within the same framework, if we assume 
that an additional real variable related to heat does exist (real in 
Einstein's sense, i.e., it can be measured). For closed systems 
(systems that do not exchange energy, volume, or matter with the 
surroundings) the new variable, the entropy $(S)$, is a homogeneous, 
first order function of the extensive controlled variables. For these 
systems, the relation $S = S(U, V, N)$ is referred to as the 
fundamental relation or, the fundamental equation. It is also assumed 
that $\it S$ is continuous and differentiable and is a monotonically 
increasing function of $U$. The assumption of the existence of the 
fundamental relation goes hand in hand with the second law of 
thermodynamics: the entropy reaches a maximum (as a function of the 
uncontrolled variables) at equilibrium.

The formulation of the second law is relatively simple, yet it is 
perhaps the most mysticism-clad law of physics. Phrases 
like ``macroscopic systems have a tendency to reach equilibrium'' or 
``the natural tendency of closed systems is to maximize their entropy''
are used freely. For example, to quote Callen \cite{Callen}: ''... in 
all systems there is a tendency to evolve towards states in which the
properties are determined by intrinsic factors and not by previously
applied external influences. Such simple terminal states are, by 
definition, time independent. They are called equilibrium states.''
The above state of affairs reminds one of Aristotelian times. Then
it was said that the ''natural state'' of bodies is to be at rest, and
that bodies have an internal tendency to reach their natural state.
The ''natural state'' has also been reflected in the terms used to 
describe the state of bodies. Non-moving objects were referred to as
bodies not at rest. We now understand that it is not an internal
tendency of bodies to be at rest. On the contrary, we need dissipation
to force the bodies to reach the minimum of a potential well and stay
there at rest. Below, we argue that in a metaphorically similar manner,
it is the gravitational background that forces the system to reach
equilibrium. 

Since its energy is controlled and the Hamiltonian describing it 
includes no interacting parts with the environment, a theoretically 
defined closed system must remain forever in one of 
its many-body quantum states. Note that this state can be either an 
energy eigenstate or a coherent superposition of energy states. In both 
cases, it is a specific microstate of the system. As entropy is a 
measure of the number of microstates corresponding to a macrostate of a 
system, it vanishes. This implies that for an ideal closed system there 
is no sense in defining and talking about the system's entropy. Usually 
this difficulty is ``solved'' by the argument that one should consider 
not a fixed value of the system's total energy $U$, but rather include 
some uncertainty so that $U$ is controlled up to some $\delta U$. 

We prefer the solution put forward by Callen \cite {Callen}:
``The apparent paradox is seated in the assumption of isolation of a 
physical system. No (finite) physical system is, or ever can be, truly 
isolated''. He mentions electromagnetic background, gravitational 
fields and the vacuum itself; all can exchange energy and matter with 
the system. A similar argument has been raised by Landau and Lifshitz
\cite{LL6}: ``In consequence of the extremely high density of levels, 
a macroscopic body in practice can never be in a strictly stationary 
state. First of all, it is clear that the value of the energy of the 
system will always be ``broadened'' by an amount of the order of the 
energy of interaction between the system and the surround bodies. The 
latter is very large in comparison with the separations between levels, 
not only for quasi-closed subsystems but also for systems which from 
any other aspect could be regarded as strictly closed. In Nature, of 
course, there are no completely closed systems, whose interaction with 
any other body is exactly zero; and whatever interaction does exist, 
even if it is so small that it does not affect other properties of the
system, will still be very large in comparison with the infinitesimal
intervals in the energy spectrum.''

In other words, Landau and Lifshitz proposed that {\it in practice} we 
cannot have ideal closed systems. We would like to argue that 
${\it in principle}$ there cannot be ideal closed systems. Consider a 
``closed'' system, which according to the argument of Landau and Lifshitz 
must be in a mixed state with some uncertainty of energy, $\delta U$. 
Except for postulating its density matrix, there is only one way to 
describe the system, which is to assume that the system and its 
environment constitute one big physical system which is prepared in a 
pure quantum state. Due to the interaction between the small system and 
the environment, the exact eigenstates of the big system are entangled. 
This means that by tracing over the states of the environment one 
obtains a reduced density matrix corresponding to the small system in a 
mixed state. The entropy of this mixed state may be calculated in the 
usual manner, and it simply reflects the measure of the entanglement with 
the environment. Within this approach one can mimic the growth of the
entropy of the small system by initially preparing the big system in 
a ``less'' entangled state. Indeed, since all the eigenstates of the 
big system are entangled, an unentangled state must be a very unique 
superposition of many eigenstates with different energies. The time 
evolution, then, will always increase the entanglement, at least for 
some initial period of time. This approach may be very useful for ``all 
practical purposes'', since the larger the environment the longer one 
can mimic the irreversibility. However, we are left with the big
system, which was prepared in a pure state. The entropy of this system
vanishes, and none of the fundamental questions is really resolved. 

We come to the conclusion that there should exist a universal 
environment (background) to which any ``closed'' system is coupled. 
This background may not be united with any physical system to form a 
bigger system in a pure state. As a result, the closed system is in a 
mixed state, entropy can be assigned to it, and it is subjected to the 
second law. The above is valid provided that the interaction with the 
background is larger than the energy spacing between the many-body 
quantum states. In other words, since a closed system is actually open,
the background induces the transitions which lead to the existence of 
entropy. Entropy can be viewed as the interaction of the system with 
its background under natural constraints (minimal interaction with the
background). The system does not have a tendency to reach 
maximum entropy; it is rather the background which forces the system 
towards equilibrium. We propose that the universal background with 
which every macroscopic system has minimal interaction is the
gravitational field. The other fields, in principle, can be either
screened or included as part of the Hamiltonian of the system.

If the microscopic states of a system are to be equally probable, so 
should be the transitions between these states. Thus the coupling with 
the background has to lead to induced transitions with equal 
probability. It might be due to the ``central limit theorem'' of random
variables when applied to the background. However, one may think about 
another possibility. The background may couple a given state of a 
system only to a small number of states (as is the case for other 
known interactions). Then a tree-like structure would be induced in the 
space of all states of the system. Starting at a given state, one may 
go only to those states which are connected to the initial one by the 
background. In the next step, another subset of states becomes 
accessible. The ``transport'' on such a tree may be very nontrivial.
In a recent work \cite{Altshuler-Gefen}, the Cayley tree structure
of states was used, and the localization on such a tree was interpreted
as a transition from the Fermi liquid picture for high energy states
to a more refined one for low energy states. It may happen that the 
localization on a background induced tree of states corresponds to 
ergodicity-nonergodicity transition. 

The essence of thermodynamics and statistical physics is that we can,
in principle, define energy and mass (numbers of particles and their
masses) for any enclosed finite volume, and decouple it from the 
environment. This implicit assumption is in contradiction with 
Einstein's theory of gravitation.  According to the latter we cannot 
co-define the energy of the gravitational field with matter in any 
enclosed finite volume. To quote Dirac \cite{Dirac}:``It is not possible 
to obtain an expression for the energy of the gravitational field 
satisfying both the conditions: (i) when added to other forms of energy 
the total energy is conserved, and (ii) the energy within a definite 
(three-dimensional) region at a certain time is independent of the 
coordinate system''. Or as Landau and Lifshitz formulate \cite{LL6}: 
`` ... the gravitational field cannot itself be included in a closed 
system, since the conservation laws which are, as we have seen, the 
foundation of statistical physics would 
then reduce to identities''. These strange properties of the 
gravitational field follow from the dual role of the metric tensor 
$g_{ij}$. On one hand, it generates the symmetry of general coordinates
transformation, i.~e., the variational derivative of the action with
respect to $g_{ij}$ is the energy-momentum tensor. On the other hand, 
as it is also a physical field (the gravitational field), the same 
derivative gives the corresponding Lagrange equation of motion. The 
energy-momentum tensor thus vanishes. We see that the gravitational 
field is the only one that cannot be screened or be included as a part
of the Hamiltonian of the system.

The above leads us to propose a new possible interpretation of the 
entropy, which can also be viewed as a postulation of a new law of 
Nature. When enclosing a volume to construct a ``closed system'', we 
impose the constraints that the system does not exchange matter, heat 
and energy besides gravitational energy with its surrounding. Thus a 
``closed system'' is ``open'' with respect to interaction with the 
background gravitational field. The latter should be viewed as an 
entropy bath (in analogy to a heat bath, particle bath, etc.), as it 
causes transitions between the system's microstates. Moreover, when the 
strength of the interaction with the gravitational field is much larger 
than the energy spacing between the many-body quantum states of the system, 
the latter becomes irrelevant, and the microstates are determined by the 
single-particle states \cite{Penrose}. In the new interpretation, the 
entropy represents the effect of the uncontrolled background on the 
enclosed system, and is not an inherent property of the system itself. 
In the same manner, the second law reflects the effect of the background 
on the system rather than being a ``tendency'' of the system.

At present we lack a theory of quantum gravity which, in our 
new picture, is necessary for the complete establishment of the 
foundation of thermodynamics and statistical physics. (It might be that 
knowing the behaviour of macroscopic systems will actually 
provide hints on the principles of quantum gravity.) Therefore, we also 
lack a quantitative evaluation of the strength of the interaction 
between the macroscopic system and the background gravitational field. 
Yet, the naive assumption is that, due to the smallness of the Planck
scale, the interaction of systems with the background gravitational field
should be neglected. Gravity is assumed to play a role either at and below
the Planck scale, or on cosmological scales. The relevant background field
for non-cosmological thermodynamic systems is assumed to be the background 
electromagnetic field which has a much stronger interaction with the system. 
Moreover, the energy of the background radiation ($3^\circ\,$K) is 
considered to be much higher than the yet unknown energy of the background 
gravitational field, as they departed from mutual equilibrium at an early 
stage of the universe \cite{Weinberg2}. Nevertheless, some estimations 
suggest that the two energies are not that different \cite{Weinberg2}. Now 
comes into play the fundamental difference between the two fields, namely, 
the fact that the electromagnetic field can be screened. Indeed we can 
perform experiments lowering the temperature of thermodynamic systems well
below $3^\circ\,$K.

Recently, the strength of the interaction with the background field has
been estimated. Ellis et al. \cite{Ellis} have proposed that the 
correction to the time evolution of the density matrix is proportional to
\begin{equation}
\label{E_Corr}
\delta E={E^2\over M_{Pl}}\ ,
\end{equation}
where $E$ is the energy of the system and $M_{Pl}=10^{19}\,$GeV is the 
Planck mass (in units of energy). In their case, the ''system'' is an
elementary or a composite particle. Adaptation of this estimate to 
thermodynamic systems can be done in two ways: 1. from the point of view 
of the individual particles composing the system. 2. from the point of 
view of the whole system. Consider a thermodynamic system of $1\,cm^3$
composed of an Avogadro number ($N_A$) of non-interacting particles at 
temperature $T=1^\circ\,$K. Taking the individual particle view, the 
energy of each particle is approximately $k_B T$ ($k_B$ is the Boltzmann
factor), hence the correction per particle, $\delta E_1$ is given by
$\delta E_1={(k_BT)^2\over M_{Pl}}\ ,$ and the s correction for the 
system is
\begin{equation}
\delta E=N_A\delta E_1={N_A (k_BT)^2\over M_{Pl}}\ .
\end{equation}
Taking the alternative interpretation, the system's total energy is 
$N_Ak_B T$, hence the correction is 
\begin{equation}
\delta E={(N_Ak_BT)^2\over M_{Pl}}\ .
\end{equation}
Inserting the parameters indicated above we obtain 
$\delta E\approx10^{-32}\,$J and $\delta E\approx10^{-9}\,$J for the first 
and second interpretation, respectively. We would like to emphasize that 
the spacing between two many-body energy levels of the system under
consideration is of the order of $10^{-40}\,$J. Thus, even if we take the 
first interpretation, the energy correction is sufficient to mix the 
energy states and lead to the emergence of entropy. As we have proposed 
before, it also leads to the breakdown of the many-body states into a 
distribution over the single-particle states (the Fermi-Dirac and 
Bose-Einstein distribution for Fermions and Bosons, respectively). The 
situation is different when the system is in a coherent macroscopic
quantum state (e.g. superfluidity, superconductivity, Hall state, etc.)
with a large energy gap separating the state from the continuum. In this
case, when the energy gap is larger than the effect of the background 
gravitational field, the latter can be ignored.

We expect that the energy correction is given by Eq. (\ref{E_Corr}) as
long as the system's energy is sufficiently high. Otherwise, there is a 
minimal correction which, in small systems, is proportional to the energy 
of the background gravitational field times the system's mass, and inversely 
proportional to the size of the system. In large systems, it is proportional 
to the time of interaction times the speed of light. The effect of the 
gravitational field should therefore saturate at a minimal level as the 
temperature of the system is lowered. This differs from the effect of the 
electromagnetic field (within the system), which decreases with the system's 
temperature and saturate only at its zero point fluctuations. Thus, the 
coupling to the background gravitational field might explain the phase 
transitions in He$^3$ \cite{Leggett} and the observations that the dephasing 
time in mesoscopic systems saturates as the temperature is lowered 
\cite{Saturation,Mohanty}. 
To explain the latter according to the new picture, one requires additional 
assumptions: 1. The dephasing due to the gravitational field is carried out
not only through direct coupling to the moving particle. The main effect is 
through the coupling of the field to the whole system which, in turn, is 
coupled to the particle via the mechanism of Stern et al \cite{SAI}. 2. The 
metric of the gravitational field is not quantized, i.e., the dephasing 
process does not require emission or absorption of gravitons. Accepting the 
above, we predict that the dephasing saturation temperature depends on the 
mass density of the system. In the experiments of Mohanty, Jariwala and Webb, 
the saturation temperature of GaAs is found to be higher, and the dephasing 
time is found to be shorter relative to those of Si. Clearly, it can also 
result from the different electronic structure of the two materials. We 
suggest to distinguish between the two possible mechanisms by using systems 
of equal dimensions made of different isotopes of the same material.

The picture above has immediate implications with respect to quantum 
measurements and the collapse of the wave function. In this picture, the
collapse is a consequence of the interaction of the particle with the
background gravitational field mediated via the measuring apparatus. We
will discuss this issue elsewhere, together with other issues related to
the new picture (e.g., the fact that the universe as a whole seems to 
evolve towards lower entropy, the evolution of complexity and entropy 
production of open systems, etc.).

To conclude, we propose that the origin of irreversibility in time is
the interaction of energy and matter with the metric of space-time
(which can be viewed as a generalized Mach-like principle), and 
that the fundamental relation of thermodynamics originates from the 
minimal interaction of any enclosed system with its gravitational 
background.

If, indeed, entropy reflects the coupling of macroscopic systems to the 
background gravitational field, the macroscopic behaviour is not simply 
derivable from the isolated microscopic dynamics of the system, and we 
may have to re-examine our reductionist view of Nature.

This article required knowledge in thermodynamics, statistical physics,
the foundation of quantum mechanics and issues related to classical and 
quantum gravity.  We were lucky to be able to learn from and consult with 
Y.~Aharonov, Y.~Bekenstein, D.~J.~Bergman and B.~Reznik, each with his 
own expertise. We also thank the referee for his constructive comments on 
the first version of the manuscript and for the valuable references he 
pointed out to us, and D.~Halbing for critical reading of the manuscript.

Note added in proof: After the submission of our manuscript we have come
across a manuscript entitled: ''Entropy Defined, Entropy Increase and 
Decoherence Understood, and Some Black-Hole Puzzles Solved'' by 
Bernard~S.~Kay \cite{Kay} which presents a similar picture from a quantum 
gravity starting point.

\end{document}